\documentclass[review]{elsarticle}
\usepackage{amsmath}
\usepackage{graphicx}
\usepackage{color}
\usepackage{hyperref}

\journal{Renewable Energy}

\newcommand\po{P1}
\newcommand\pt{P2}
\newcommand\podg{\po$_{\text{DG}}$}
\newcommand\popt{\podg$-$\pt}

\newcommand\pp[2]{\frac{\partial #1}{\partial #2}}
\newcommand\ppt[1]{\pp{#1}t}
\newcommand\dd[2]{\frac{d #1}{d #2}}
\newcommand\ddt[1]{\dd{#1}t}
\newcommand\grad\nabla
\renewcommand\div{\nabla\cdot}

\begin{document}
\begin{frontmatter}
\title{A correction to the enhanced bottom drag parameterisation of tidal
  turbines}
\author[amcg]{Stephan C. Kramer\corref{mycorrespondingauthor}}
\cortext[mycorrespondingauthor]{Corresponding author}
\ead{s.kramer@imperial.ac.uk}
\author[amcg,grantham]{Matthew D. Piggott}
\address[amcg]{Applied Modelling and Computation Group,\\
Department of Earth Science and Engineering,\\
Imperial College London,\\
South Kensington Campus, London SW7 2AZ,\\
United Kingdom}
\address[grantham]{Grantham Institute for Climate Change and the Environment,\\
  Imperial College London, United Kingdom}
\begin{abstract}
Hydrodynamic modelling is an important tool for the development of tidal stream
energy projects. Many hydrodynamic models incorporate the effect of tidal
turbines through an enhanced bottom drag. In this paper we show that although
for coarse grid resolutions (kilometre scale) the resulting force exerted on
the flow agrees well with the theoretical value, the force starts decreasing
with decreasing grid sizes when these become smaller than the length scale of
the wake recovery.  This is because the assumption that the upstream velocity
can be approximated by the local model velocity, is no longer valid. Using
linear momentum actuator disc theory however, we derive a relationship between
these two velocities and formulate a correction to the enhanced bottom drag
formulation that consistently applies a force that remains closed to the
theoretical value, for all grid sizes down to the turbine scale. In addition, a
better understanding of the relation between the model, upstream, and actual
turbine velocity, as predicted by actuator disc theory, leads to an improved
estimate of the usefully extractable energy. We show how the corrections can be
applied (demonstrated here for the models MIKE 21 and Fluidity) by a simple
modification of the drag coefficient.
\end{abstract}


\end{frontmatter}

\section{Introduction}
One of the key advantages of tidal energy as a renewable energy source, 
is the predictable nature of the resource. Methods for 
the detailed prediction of tidal dynamics using hydrodynamic numerical 
models have been developed over many years and have been applied for many
different purposes. Less well understood is how the placement of 
tidal energy converters in the flow will modify the existing tidal
currents at both local and regional scales \citep{adcock15}. The challenge here is that the detailed flow 
around a turbine is a three-dimensional phenomenon comprising far smaller length scales
than those of the underlying tidal resource. A typical approach therefore is 
to model the turbine scale flow in a three-dimensional CFD simulation based on a actuator disc,
blade element, or actuator-line model (see e.g.
\citet{sun08,harrison10,batten13,malki13,churchfield13}). 
  The effects of the turbine in a
large scale hydrodynamic model are then parameterised, based on properties
extracted from the CFD model.

The main property of the turbine that needs to be parameterised
is the amount of thrust force exerted by the turbine on the 
flow (and vice-versa)
 as a function of the flow speed. This also determines the 
amount of energy taken out of the flow. Thrust curves typically
take the form of a quadratic function of current speed with a non-dimensional
thrust coefficient, and can be derived as described above in a high-resolution CFD model, or in
lab experiments.
Turbine specific
properties such as cut-in and rated speeds however, mean that 
the curve does not necessarily follow a quadratic and therefore
the coefficient is not constant but itself varies as a function 
of flow speed.

It is important to note, that the turbine properties 
derived in e.g. a CFD model, or from lab experiments, typically consider the placing of a single
turbine in uniform background flow. Speed dependent properties are then expressed
in terms of the background velocity, which, because the velocity is 
slowed down in the presence of a turbine, is available as the undisturbed upstream velocity.
In a finite width channel, blockage effects may also affect the resulting thrust
curve but can be corrected for (see e.g. \citet{garrett07,whelan09}) to derive the thrust curve for
an idealised free-standing turbine.  In addition, the results
may be dependent on the turbulent properties of the flow.

An approach followed in many models
is to implement the thrust in the form of an equivalent
drag force term. For depth-averaged models this effectively comes down to an increased bottom
drag \citep{lalander09, vennell10,plew13,funke14,martin15}
  Three-dimensional models may implement the drag as a force over the entire
water column \citep{defne11}, or if the vertical resolution allows it the drag can be applied over a
vertical cross section (e.g. \citet{roc13}), i.e. an idealised actuator disc.  

Since the thrust force is given as a function of the upstream velocity, it is
important to consider what velocity to use for the equivalent drag force in the
model.
One option is to probe the
numerical velocity solution somewhat upstream of the turbine 
location. This however brings with it various difficulties such as the question
of how far upstream is appropriate, or the fact that the flow upstream 
might not actually return to the uniform background flow condition that was
considered in the CFD model, due to bathymetric changes or the presence
of other turbines. Additionally, the use of a non-local velocity is 
not desirable for numerical and computational purposes: it makes it hard 
to treat the term implicitly (in the time-integration sense), potentially leading to
time step restrictions for stability, and memory access outside 
of a fixed numerical stencil, or across sub-domains in domain-decomposed parallel
model, is computationally inefficient.

When enough mesh resolution is available, both in the horizontal and vertical
dimensions, to resolve the flow through the turbine the relationship between the
upstream velocity and the turbine velocity
 can be predicted using 
Linear Momentum Actuator Disc Theory (LMADT). Using this relationship the quadratic drag law
can be reformulated into a function of the local velocity, thus overcoming
the difficulties and ambiguities mentioned above. This is the approach followed
in \citet{roc13}. The typical width of a tidal turbine, order 20m, can however
be orders of magnitude smaller than the spatial scales of the tidal flow so 
that resolving an individual turbine may become prohibitively expensive
even in large-scale unstructured mesh models that allow for the efficient focusing of 
mesh resolution. Also this approach requires the alignment of the mesh with
the position and direction of the turbine, thus limiting the flexibility 
to quickly evaluate different turbine positionings and angles.

If the mesh resolution available is such that computational cells are much larger
than the turbine scale, the drag force is necessarily applied over a larger
area. In a typical implementation a constant drag is applied over 
a single cell (the cell that contains the turbine). If the cell size
is in fact large enough it may be expected (this
will be further investigated in this paper), that the local velocity is
not actually affected greatly by the presence of the drag term since the
drag force is ``smeared'' out over a large area and the local cell velocity
represents an average of the velocity in a large area around the 
turbine. In that case the difference between the undisturbed background
flow and the local cell velocity may be neglected and the turbine 
can be implemented using a simple quadratic drag law, function of the local velocity.

As will be shown in this paper however, when the mesh resolution is refined
such that mesh distances are closer to the turbine scale, this approximation
is no longer tenable as the difference between upstream and local velocity 
becomes too large. As long as individual turbines are not resolved however,
the approach in \citet{roc13} is also not valid as the local velocity
is still larger than the theoretical turbine velocity predicted by
linear momentum actuator disc theory. In particular, for depth-averaged models
the local velocity will remain higher than the actual turbine 
velocity even when the horizontal scales are sufficiently resolved. This is due
to the fact that the drag acts on the entire water column and thus the
depth-averaged model velocity will represent an average of the actual turbine
velocity and a higher by-pass velocity above and below the turbine. Even in
three-dimensional models the drag force is often applied over the entire water
column \citep{defne11}, or limited to one or only a few
layers \citep{yang13,hasegawa11}, and does not
necessarily give an accurate representation of the actual turbine cross-section
and thus the model velocity where the drag is applied is not necessarily equal
to the real turbine velocity.

Here we demonstrate how the actuator 
disc computation may be modified to include the fact that the drag force
numerically is applied over a different cross section than the actual turbine. Thus again
an analytical relationship can be derived between the undisturbed 
upstream flow and the local cell velocity, and similarly the drag force can
be reformulated as a drag law dependent on the local cell velocity. Like
the approach in \citet{roc13}, this leads to a correction to the drag 
law, which in this case depends on the local cell width, but that nonetheless
can easily be implemented in existing models,  as will be demonstrated here for
the Fluidity and MIKE 21 models.

An alternative method for the parameterisation of turbines in large-scale
hydrodynamic models that also makes extensive use of actuator disc theory,
is the line momentum sink method \citep{draper10,serhadlioglu13}. Actuator disc
theory is used to express the effect of turbines, and more specifically 
an entire fence of turbines, as a relative head loss across
the whole near-field flow pattern starting from the assumed uniform upstream flow at one
end , down to the end of individual turbine wakes at the point where uniform flow is
again achieved (within the far-field wake of the fence). This head loss is then
applied as a jump condition across an edge, or multiple aligned edges within the
computational grid using a Discontinuous Galerkin discretisation of
the depth-averaged shallow water equations. The advantage of this method that it
incorporates a detailed LMADT treatment of the near-field effects, including
blockage effects for multiple turbines in a fence. It does require however that
these effects are treated at the sub-grid level, and is therefore only
appropriate for hydrodynamic models with grid sizes larger than the length scale
of the near-field/turbine wake (typically 10--20 turbine diameters)
\citep{draper10}.

For any numerical modelling study it is important to look at the effect of
changing the grid resolution on the results of interest. In the modelling
guidelines for tidal resource assessments in \citep{legrand09}, a range of
grid resolutions is recommended depending on the stage of the resource
assessment, ranging from kilometre scale for regional studies, down to a range of
500~m to 50~m for specific site feasibility studies.
Since the wake of a turbine is a three-dimensional phenomenon, it is not
expected that an accurate description of the near-field flow can be obtained
with a depth-averaged model. Nevertheless, such models should be capable of
studying far-field effects. This relies on the correct forces and their effect
on the large-scale flow being modelled correctly. As this paper shows however,
the results of the standard enhanced bottom drag parameterisation of 
the turbine thrust force will deteriorate as 
the mesh resolution falls below that of the near-field/wake length
scale ($\approx 200-300$ m for a typical turbine). The correction
proposed in this paper ensures that consistent results can be obtained with grid
resolutions smaller than the length scale of the turbine wake, all the way down to
the turbine scale.

\section{Enhanced bottom drag formulation}
\label{sec:enhanched_bottom_drag}
In this section we will describe the enhanced bottom drag parameterisation of turbines
used in many models \citep{plew13,divett13,martin15,yang13} and demonstrate some issues with mesh dependency.
We will do this within the framework of MIKE 21 \citep{mike}, a depth-averaged
hydrodynamics model widely used in the marine renewable industry, and an equivalent
drag-based implementation in Fluidity, an open source, finite element modelling
package \citep{piggott08,fluidity15}. By
comparing results between the two models 
we verify that the implementation in the closed source model MIKE 21 is indeed 
based on the same theory that underlies our implementation in Fluidity, and
that the same issues are observed.

The aim of the turbine parameterisation is to represent the drag force of the
turbine on the flow, which is typically given as:
\begin{equation}
  \vec F(\vec u) = \tfrac 12\rho C_t(|\vec u|) A_t |\vec u|\vec u,
  \label{eq:quadratic_force}
\end{equation}
here $\vec u$ is the flow velocity, $\rho$ the density of sea water, $C_t$ the
dimensionless drag or thrust coefficient, and $A_t$ the effective cross-sectional
area of the turbine in the flow. The drag coefficient $C_t$ may itself be a function
of speed due to turbine properties such as rating, pitch control and the use of
a cut-in speed.
As discussed in the introduction the drag law, often derived from
a small-scale three-dimensional CFD model, is typically expressed as a function
of the undisturbed
background flow velocity, which in the case of an idealised domain corresponds
to the uniform velocity upstream of the turbine.

The depth-integrated  shallow water equations (in conservation form) are given by
\begin{align}
  \ppt{H\vec u} + \div \left( H\vec u \otimes \vec u\right)
  + gH\grad\eta + c_b |\vec u|\vec u &= 0, \label{swe_momeqn} \\
  \ppt\eta + \div \left(H\vec u\right) &= 0,
\end{align}
where $H$ is the total water depth between bottom and free surface, elevated
at a level $z=\eta$, $\vec u$ is the depth-averaged velocity, $g$ the
gravitational acceleration and $c_b$ is the bottom friction coefficient.

A local
momentum balance in a fixed local horizontal area $A$ is derived by integrating
\eqref{swe_momeqn} over this area, multiplied by $\rho$:
\begin{equation}
  \ddt{} \int_A \rho H\vec u 
  + \int_{\partial A} \rho H \left(\vec n\cdot \vec
  u\right)\vec u 
  + \int_A \rho gH\nabla\eta
  + \int_A c_b\rho|\vec u|\vec u = 0.
\end{equation}
The second term represents momentum flux through the boundary $\partial A$.
The third term can be rewritten as an integral of hydrostatic pressure around
the three-dimensional water column below $A$.
 The last term represents a momentum sink term due to
bottom friction.

To implement the turbine thrust force through an enhanced
bottom friction, $c_b \to c_b+c_t$, 
we need the additional momentum sink to be equal to the force, $\vec F(\vec u)$
in \eqref{eq:quadratic_force}.
To address the question of which velocity $\vec u$
is used to compute $\vec F(\vec u)$, in a first attempt we simply employ the
local, depth-averaged
velocity and average the force over the area $A$. Thus, we require that
\begin{equation}
  \int_A c_t\rho|\vec u|\vec u = \frac{\int_A \vec F(\vec u)}A.
  \label{eq:force_averaged}
\end{equation}
Combined with \eqref{eq:quadratic_force}
, it readily follows that the enhanced bottom drag
coefficient $c_t$ in this case should be set to:
\begin{equation}
  c_t(\vec u) = \frac{C_t(\vec u) A_t}{2A}.
  \label{eq:enhanced_drag_coef}
\end{equation}

Since we consider the parameterisation of turbines in hydrodynamic models where
mesh distances are larger than the size of an individual turbine,
the force is applied over the smallest area possible, typically the area
of a single mesh cell.
Thus the area $A$ in
\eqref{eq:enhanced_drag_coef} corresponds to the cell area over which the
enhanced drag coefficient is applied.
In models
where the cell area is much larger than the turbine cross section $A_t$, the
additional drag is small and therefore the presence of the turbine will
not have a large effect on the numerical solution for $\vec u$ in that cell.
As an example, for typical values of $C_t=0.6$, a mesh distance $\Delta
x=200~m$ and turbine diameter $D=18~m$, if the drag is applied over a single
square computational cell of $\Delta x\times\Delta x$, we get
\begin{equation}
  c_t(\vec u) = \frac{C_t \pi \left(\frac D2\right)^2}{2 \Delta x^2} \approx
  0.00122,
\end{equation}
which is only half of a typical value of $c_b=0.0025$ for the background bottom friction coefficient.

Since the effect of the additional drag is relatively small it is to be expected that
the assumption that the local velocity within the cell is close to the
undisturbed background flow is valid for relatively coarse resolution models, and can
therefore be used in the averaged force in the right-hand side of
\eqref{eq:force_averaged}.
As the resolution is
increased however and the mesh distances become closer to the turbine scale, 
the drag is applied over a smaller area and the reduction
in local flow speed may become much larger. Because of the 
quadratic dependency of the drag force on the flow speed, this may have a
significant impact on the force that is applied in the model.

\section{Local velocity drop in idealised channel}
We investigate the mesh-dependent reduction in local flow speed in more
detail in the following
idealised set up: a turbine is placed in a rectangular channel of length
10 km and width 1 km. The depth at rest is set to 25m and a bottom
friction of $c_b=0.0025$, equivalent to a Ch\'ezy coefficient of $62.6$
m$^{1/2}$s$^{-1}$, is applied. At the upstream
boundary a uniform velocity of 3.0 ms$^{-1}$ is enforced. At the downstream
end a Flather boundary condition is applied. The steady state solution
without a turbine can be described as a balance between the free surface
gradient and the bottom friction. The necessary free surface slope leads to a water
level that is approximately $0.9$ m higher at the upstream boundary than at the
downstream boundary. The decrease in
water depth $H$ along the channel, in combination with the continuity equation,
leads to an acceleration along the channel, with the speed increasing from $3.0$
ms$^{-1}$ to
$\approx 3.12$ ms$^{-1}$ downstream. In a separate computation of the
hydrodynamics without a turbine, it was
verified in both Fluidity and MIKE 21, that the background flow velocity at the turbine location,
halfway the channel, is approximately 3.055 ms$^{-1}$.

For the simulations with a turbine, the following turbine parameters were chosen: the
thrust coefficient $C_t=0.6$ with a turbine diameter of $D=16$ m giving a turbine
cross-sectional area of $A_t=201$ m$^2$. The simulations were performed using both MIKE 21 and
Fluidity on a series of identical triangular meshes with uniform resolutions starting at
a mesh size of $\Delta x=320$ m, doubling the resolution each time with
the mesh size decreasing down to $\Delta x=20$ m. One extra, fine resolution 
mesh with a mesh size equal to the turbine diameter was then run, $\Delta x=D=16$
m. For the parameterisation of the turbine in
Fluidity the enhanced bottom drag approach described in the previous section
was chosen. Although Fluidity here uses a finite element scheme with
discontinuous piecewise linear velocity and piecewise quadratic
pressure solution (the mixed \popt\ velocity--pressure element pair, see
\citet{cotter09}) a
piecewise constant drag field was used to simplify the computations and to
remain close to the numerics of MIKE which uses a finite volume scheme with
higher order flux reconstructions. Although the exact details of the
implementation in MIKE were not available, the results between Fluidity and MIKE
were found to be close enough to extend the analysis based on the 
parameterisation used in Fluidity to that in MIKE 21.

\begin{figure}
  \includegraphics[width=\textwidth]{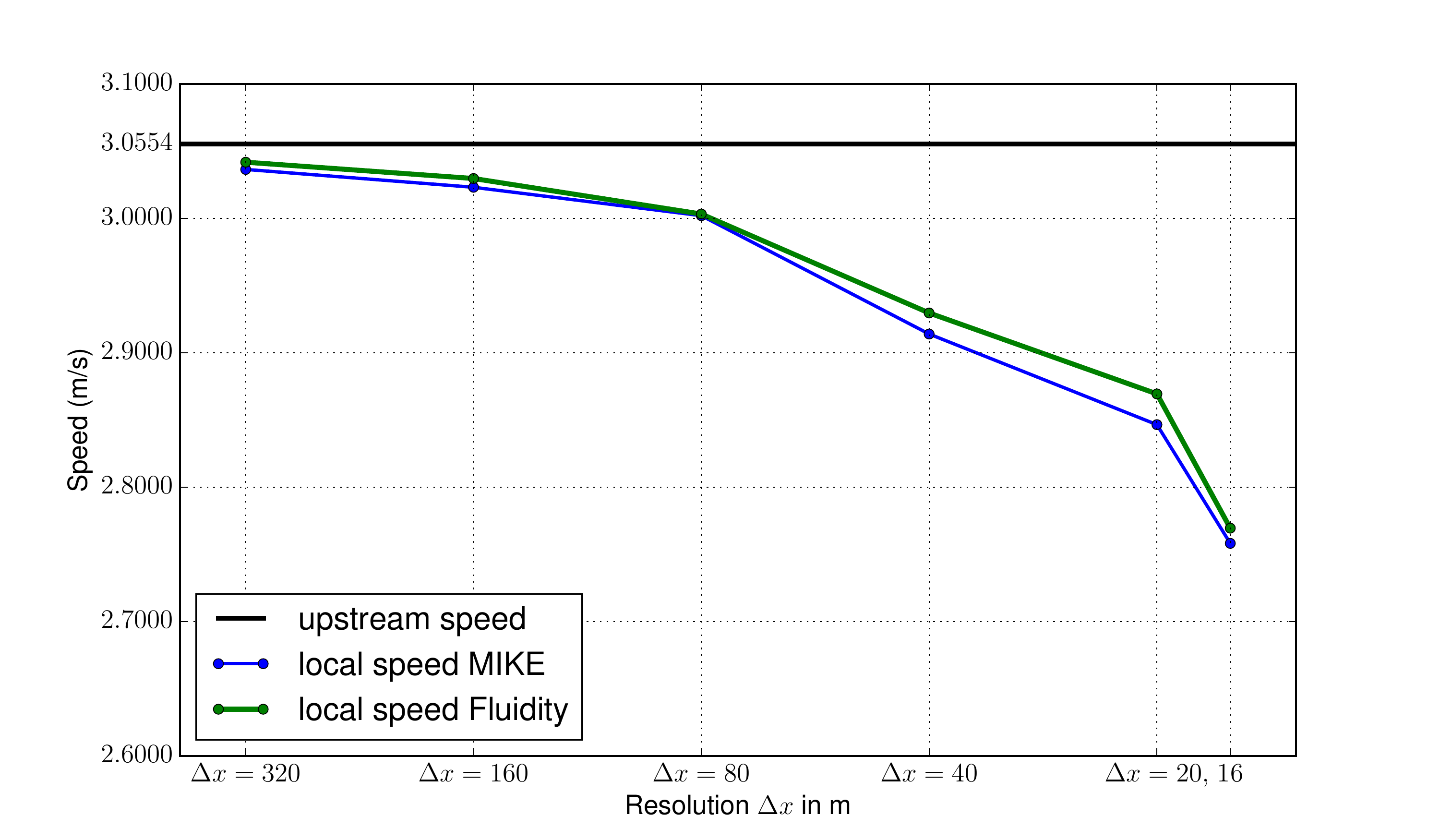}
  \caption{The speed at the turbine location, inside the enhanced bottom
    drag cell, decreases with increasing resolution both in the Fluidity and
    MIKE results.}
  \label{fig:decreasing_figure}
\end{figure}
Figure \ref{fig:decreasing_figure} displays the obtained velocity in 
the cell in which the drag has been enhanced to parameterise the effect of a
turbine. MIKE 21 employs a cell centred scheme, so for this model we report the
value in the centre of the cell. In Fluidity's numerical scheme, \popt, the
velocity is represented by a linear function in each cell which is discontinuous
between the cells. Here, and in the rest of the paper, the presented results for the local
cell velocity are obtained by taking the cell average. From figure
\ref{fig:decreasing_figure} it can be
seen that the obtained velocity is indeed highly mesh-dependent, and drops with increased
mesh resolution. 
Since the square of this velocity is used to implement the
drag term, a 10\% drop in the
local velocity leads to an approximate 20\% drop in the drag force.

In a model of a fully resolved turbine the local velocity is expected to drop.
After all,
the velocity through the turbine is known to be smaller due to momentum
exchange with the turbine, whereas the bypass flow around the turbine is
expected to accelerate. The deceleration of the flow through the turbine can
be estimated using linear momentum actuator disc theory (LMADT, see
\cite{garrett07} for an application of this theory to tidal turbines). The theory
assumes inviscid flow and a uniform upstream velocity $u_0$. Furthermore, it defines a velocity $u_1$ through the turbine,
and velocities $u_3$ and $u_4$ in respectively the wake and bypass flow (see
figure \ref{fig:square_theory}).
 It
also defines pressures: $p_0$ for the upstream pressure, $p_1$ and $p_2$
directly on either side of the turbine, and a uniform pressure $p_4$ downstream
where the velocities $u_3$ and $u_4$ are defined. At the same downstream
location, the cross-sectional area of the wake flow is defined as $A_3$.
In addition, it defines the known cross sections $A_c$ for the total channel cross section
and $A_t$ for the turbine cross section.

Through selective application of the
continuity equation, momentum conservation and Bernoulli's principle, seven
equations can be derived for the unknowns $u_1, u_3, u_4, p_1, p_2, p_0$ and $A_3$,
given $u_0$ and $p_4$ as upstream and downstream boundary conditions
respectively (see \ref{sec:lmadt}). These equations can be simplified greatly
by assuming $A_t \ll A_c$, which means no blockage effects are taken into
account. For this case, $u_4=u_0, p_4=p_0$ and 
the velocity through the turbine can be computed as (cf. equation
\eqref{u1fromu0} in the appendix):
\begin{equation}
  u_1 = \tfrac 12 \left(1+\sqrt{1-C_t}\right) u_0.
  \label{eq:local_velocity}
\end{equation}

For our idealised channel case considered above, we may compute $u_1=2.49$
ms$^{-1}$. As we will see however in the next section, the difference between
upstream and turbine velocity is smaller in the numerical results because the
drag force is spread out over an area with a larger width. As already discussed,
this means that at very coarse
mesh resolutions the velocity in the drag cell is hardly different from the
upstream velocity. As the mesh resolution is refined however, and the force can
thus be applied over a smaller width, this velocity will drop. This decrease in
local velocity will continue with increasing mesh resolution, and only when the
resolution is sufficient that the drag force can be applied numerically over 
exactly the same cross section as that of the turbine, e.g. in a
three-dimensional model, should we
expect this velocity to have reached the value of $u_1$ computed from standard
actuator disc theory. 

\section{Predicting the reduced velocity in the enhanced drag cell}
\label{sec:predict_velocity_drop}
\begin{figure}
  \def\svgwidth{\textwidth}
  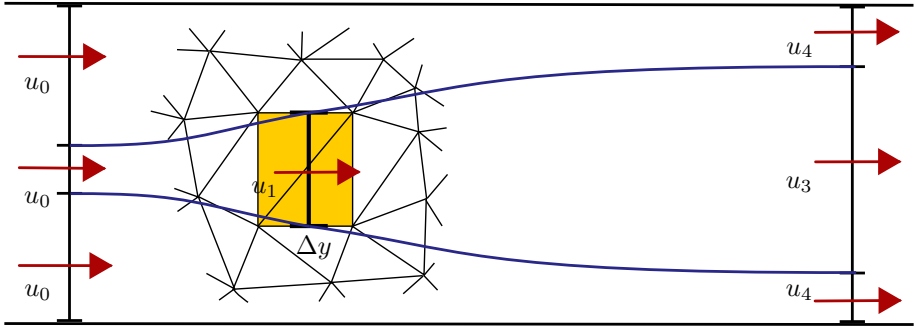
  \caption{Approximation of the enhanced drag formulation by actuator disc
    theory. An upstream velocity $u_0$ is assumed to reduce to a ``turbine''
    velocity $u_1$ inside the square in which the enhanced drag is applied. The
    effect of the enhanced drag is assumed to be equivalent to an actuator disc
    of width $\Delta y$, the width of the cell in the direction transverse to
    the flow. The relationship between $u_1$ and $u_0$ can be estimated using
    actuator disc theory which involves eliminating wake and bypass
    velocities $u_3$ and $u_4$ from a set of algebraic equations derived from
    selectively applying mass and momentum conservation and Bernoulli principles
    (see \ref{sec:lmadt}).}
  \label{fig:square_theory}
\end{figure}
To simplify matters, we first consider the case where the turbine
is represented by a square area of enhanced drag instead of a triangle. 
Additionally, we assume that this square area is aligned with the flow.
To this
end we create a series of meshes with the same resolutions $\Delta x=320$ m to
$\Delta x=16$ m as in the previous section, but with an embedded square centred
around the turbine location, of dimensions $\Delta x\times\Delta x$. The square 
is divided into two triangles, and outside the square an unstructured
triangular mesh of approximately uniform resolution is created, as indicated in figure \ref{fig:square_theory}.

When we neglect variations in the streamwise-direction (here denoted as
the $x$-direction), the model results should correspond to those of an
infinitely thin actuator disc as is considered in actuator disc theory. The
actuator disc modelled by this shallow water model has a cross-sectional area of
$\Delta y H$. Here, and in the rest of the paper, $\Delta y$ is the width of 
the drag area, in the cross-stream direction. In this section in particular
$\Delta y=\Delta x$. Since we consider mesh resolutions where $\Delta y>D$,
and additionally $H>D$, this
``numerical'' cross section will be much larger than the actual cross section
$A_t$. Therefore, to predict the results of the shallow water model using
actuator disc theory, we should be careful to use the cross section $\Delta y H$ applicable to this model.
If we neglect any variations within the horizontal square, we may then hope to predict the
velocity within the square as the disc velocity $u_1$ from this modified
actuator disc theory calculation.

Following the assumption made above \eqref{eq:force_averaged}, the magnitude of the force applied in the enhanced bottom drag approximation
is given by:
\begin{equation}
  F = \tfrac 12\rho A_t C_t u_1^2.
  \label{eq:force_from_u1}
\end{equation}
Note that here we need to use the actual turbine cross section $A_t$ as that is
the user input in this formulation to calculate the enhanced drag $c_t$ in
\eqref{eq:enhanced_drag_coef}. Further we assume that the velocity that is used to
compute the force in this approximation, which is simply the local velocity in
the drag cell, will be accurately predicted as the velocity $u_1$ in the modified actuator disc theory that follows below.

Following the steps in the derivation of \eqref{eq:local_velocity},
\eqref{u1fromu0} in the appendix, but
now applied to an actuator disc of cross section $\hat{A}_t = \Delta y H$, we 
first define
a modified thrust coefficient (cf. \eqref{ct_definition} in the appendix):
\begin{equation}
  \hat{C}_t := \frac F{\tfrac 12\rho\hat{A}_t u_0^2}
  = \frac{A_t}{\hat{A}_t}\frac{u_1^2}{u_0^2}C_t.
  \label{eq:modified_thrust_coefficient}
\end{equation}
Following the same derivation of \eqref{eq:local_velocity}, we then obtain
a relationship between the local model velocity $u_1$ and the upstream velocity
$u_0$ if in \eqref{eq:local_velocity} we replace $C_t$ with $\hat{C}_t$.
This gives an expression for the ratio $u_1/u_0$ than can be substituted in
\eqref{eq:modified_thrust_coefficient}, to give:
\begin{equation}
  \hat{C}_t = \frac{A_t}{\hat{A}_t}\left(\tfrac
    12\left(1+\sqrt{1-\hat{C}_t}\right)\right)^2 C_t.
\end{equation}
After some algebraic manipulation\footnote{the authors made use of SymPy, a
  python library for symbolic mathematics: \url{www.sympy.org}}, this can be reworked to
\begin{equation}
  \hat{C}_t = \frac{\frac{A_t}{\hat{A}_t}C_t}{\left(1+\tfrac
      14\frac{A_t}{\hat{A}_t}C_t\right)^2}.
\end{equation}
Finally, the relationship between the local velocity $u_1$ within the cell that the
enhanced drag is applied in, and the upstream velocity $u_0$ is given by
\begin{equation}
  u_1 = \frac 1{1+\tfrac 14\frac{A_t}{\hat{A}_t}C_t} u_0.
  \label{eq:u1_predict}
\end{equation}

\begin{figure}
  \includegraphics[width=\textwidth]{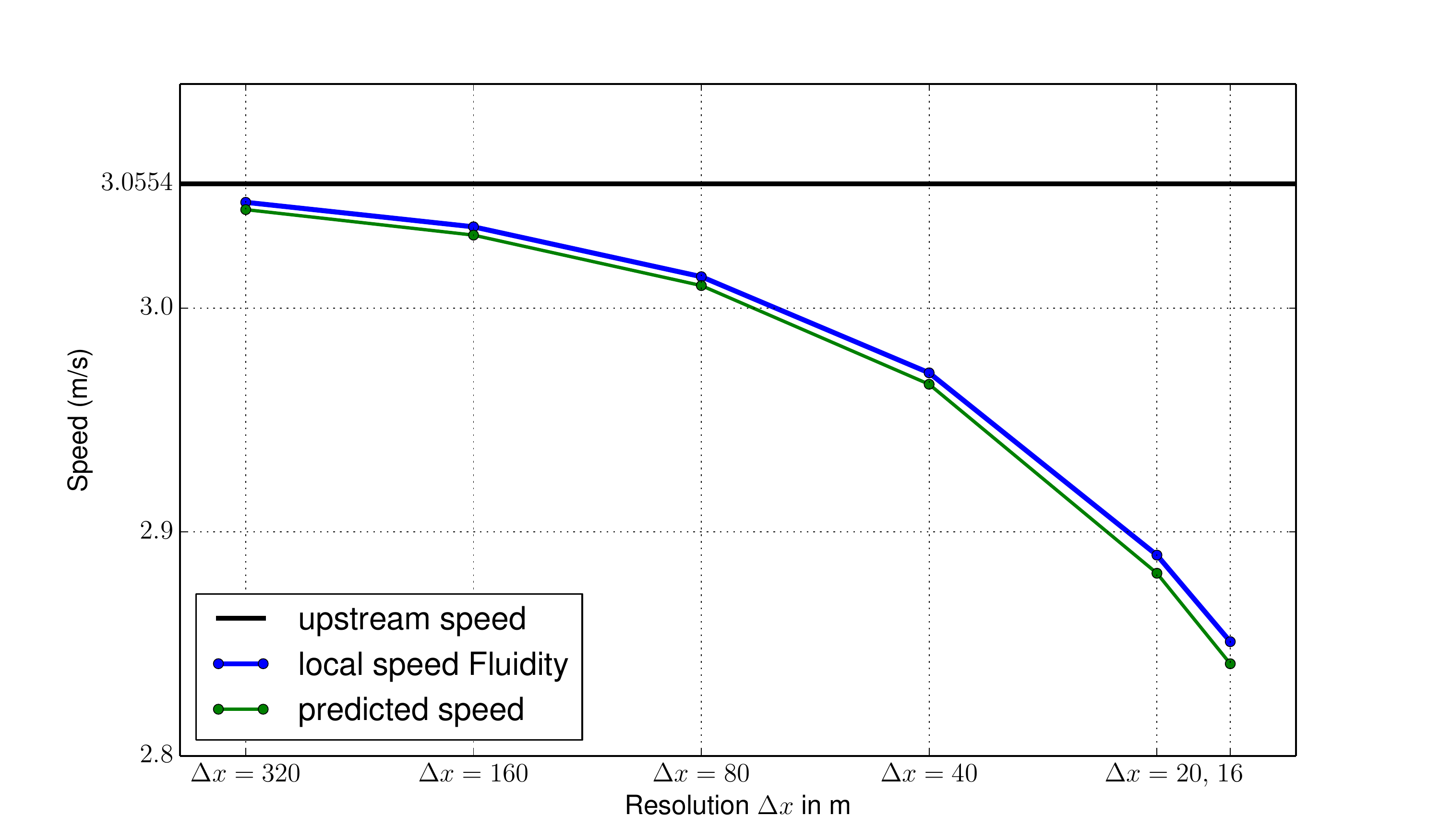}
  \caption{The speed inside a square drag cell decreasing 
    with  increasing resolution. Results are model outputs
    from Fluidity. The plotted speed is the average value
    over the square area. The
    decreasing cell speed can be accurately predicted using
    \eqref{eq:u1_predict} derived from 
    actuator disc theory.}
  \label{fig:decreasing_figure_square}
\end{figure}

Figure \ref{fig:decreasing_figure_square} shows that the speed predicted by
\eqref{eq:u1_predict}
closely follows that computed with Fluidity. Note that the results here differ
from the Fluidity results in figure \ref{fig:decreasing_figure}. This 
is because in figure \ref{fig:decreasing_figure}, the drag is applied over an
arbitrary triangle in an unstructured, triangular mesh generated by the mesh
generator with a characteristic edge length set to the value of $\Delta x$ on
the $x$-axis. The meshes used for the results here, figure
\ref{fig:decreasing_figure}, are also unstructured,
triangular and use the same characteristic edge lengths, but incorporate a
square, consisting of two triangles, with dimensions $\Delta x\times\Delta x$ over
which the drag is applied. Comparing the two figures, it can be observed that the drag being applied over a square area, aligned
with the flow direction, leads to a different relationship between the upstream
velocity and the velocity within the drag area, than when the drag is applied
over a triangle. For this reason, in the following we will derive two different
corrections to the enhanced bottom drag formulation for these two cases.

\section{Turbine correction for square cells}
\label{sec:square_cells}
We have shown that actuator disc theory, using the width of a square
enhanced drag cell and water depth, can accurately predict the relationship between upstream
and local cell velocities. This can be used to reformulate the drag force applied
in the cell to be a function not of the local cell velocity, but effectively of the upstream
velocity. Instead of applying the force in \eqref{eq:force_from_u1}, which is
based on neglecting the difference between upstream and local velocity,
we want to apply the force
\begin{equation}
  F = \tfrac 12\rho A_t C_t u_0^2,
  \label{eq:force_from_u0}
\end{equation}
where $u_0$ is the upstream velocity which is not readily (and locally)
available. This expression is the same as in standard actuator disc theory,
except now we need to take into account that this force is not 
applied over the cross-section $A_t$ but over a cross-section
$\hat{A}_t=\Delta y H$. Thus we obtain a modified thrust coefficient:
\begin{equation}
  \hat{C}_t := \frac F{\tfrac 12\rho\hat{A}_t u_0^2} = \frac{A_t}{\hat{A}_t}
  C_t.
\end{equation}
Note that this modified thrust coefficient differs from the one in the previous section, used to predict the results in the unmodified
enhanced drag formulation, as we now assume that the correct force is
applied.

Assuming $u_1$ is an adequate estimate for the local velocity in the cell with
enhanced drag $c_t$ and cell area $A$, the force applied by the enhanced drag is
given by (cf. the left-hand side of \eqref{eq:force_averaged}):
\begin{equation}
  F = \rho A c_t u_1^2.
\end{equation}
After updating \eqref{eq:local_velocity} to use the modified thrust coefficient
$\hat{C}_t$, we can substitute it here to 
make $F$ a function of the upstream velocity $u_0$:
\begin{equation}
  F = \rho A c_t\tfrac 14\left(1+\sqrt{1-\hat{C}_t}\right)^2 u_0^2.
\end{equation}
To obtain the appropriate value of $c_t$ we simply equate this expression
with the desired force in  \eqref{eq:force_from_u0}. This leads to:
\begin{equation}
  c_t = \frac{C_t A_t}{2A} \frac{4}{\left(1+\sqrt{1-\frac{A_t}{\hat{A}_t}C_t}\right)^2}.
  \label{eq:drag_correction_square}
\end{equation}
In comparison with \eqref{eq:enhanced_drag_coef} from the standard
enhanced bottom drag formulation, we have obtained an additional factor that
corrects for the fact that we are using the local cell velocity instead of the upstream velocity. For coarse
resolution runs, we have $A_t/\hat{A}_t\to 0$, and thus we fall back, as expected, to the unmodified enhanced drag formulation,
since the cell velocity is close to the upstream velocity.
 As we have seen for finer
resolutions, still coarser than the turbine scale, the difference between cell
and upstream velocities becomes significant.

The correction derived above can also be applied to three-dimensional simulations 
with a resolved turbine, where the drag force is applied in three-dimensions over a vertical
cross-sectional area (actuator disc) with $\hat{A}_t=A_t$ and
therefore $\hat{C}_t=C_t$.
The correction
factor then simplifies to exactly that given in \citet{roc13}. 
For the unresolved case however, both in two and three dimensions, the correction derived here not only corrects for the difference between upstream
and turbine velocity, but also for the difference between the actual turbine
cross-section and the cross-section over which the drag is applied numerically.

\begin{figure}
  \includegraphics[width=\textwidth=0.6\textwidth]{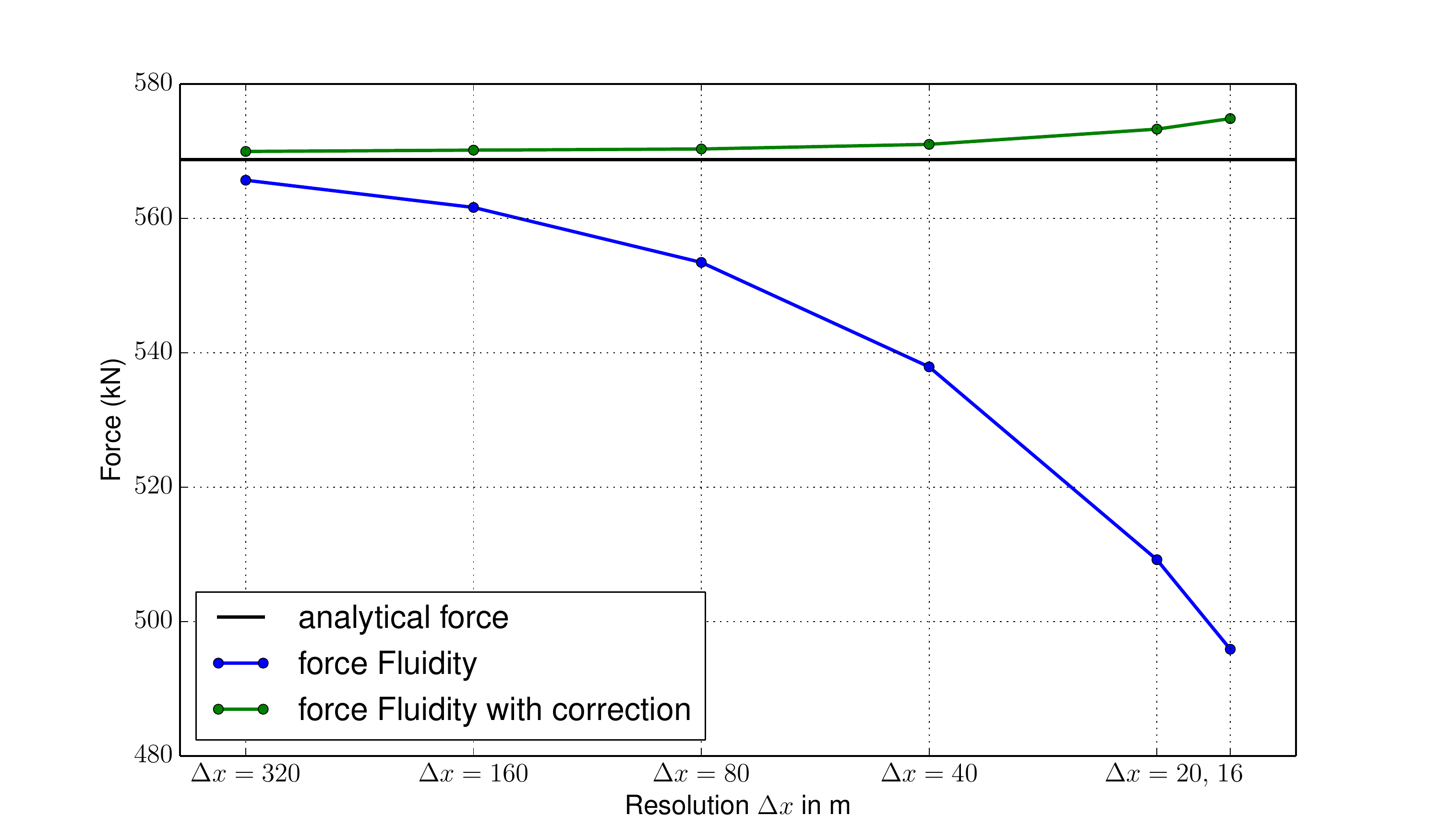}
  \caption{In the standard enhanced drag formulation for tidal turbines,
    equation \eqref{eq:enhanced_drag_coef}, the
    applied force is a quadratic function of the local velocity in the drag
    cell (here, a square area of $\Delta x\times\Delta x$). As the mesh resolution 
    increases, the local velocity drops, and
    therefore the force that is applied within the model decreases. Using the
    correction in \eqref{eq:drag_correction_square} however, the same force can
    be maintained more or less independent of resolution.}
    \label{fig:decreasing_force_square}
\end{figure}

Returning to our idealised channel case, in figure
\ref{fig:decreasing_force_square} it
is shown how the force in the standard enhanced bottom drag formulation
applied to a square decreases with increasing mesh resolution. It is to be noted 
that the relative drop in drag force is larger than the relative drop in speed,
due to the quadratic dependency of the force on the speed. Adjusting the drag
formulation according to \eqref{eq:drag_correction_square}, the applied force
is not only more accurate at coarse resolution, but also 
remains much closer to that computed from the upstream velocity directly
as the mesh resolution approaches the turbine scale.

\newcommand\dy{\mathrm{d}y}
\newcommand\dF{\mathrm{d}F}
\section{Turbine correction for triangular cells}
\label{sec:triangular_cells}
We now return to the case where the enhanced drag formulation is applied
to a single triangular cell, not necessarily aligned in any way with the flow.
Again we may approximate the applied drag by an
actuator disc spanning the width of the triangle. In this case however,
 if we thus collapse the applied drag force to a single line, the amount
 of drag varies along the disc.

We assume here that the streamlines run parallel through the triangle and use a local
coordinate system where $x$ is in the streamwise direction and
$0\leq y\leq\Delta y$ in the
transversal direction, where $\Delta y$ is the largest width of the triangle.
We may subdivide the triangle into a number of streamtubes
of infinitesimal width $\dy$, which can be considered as rectangles $\Delta
x\times \dy$, whose length
$\Delta x = \Delta x(y)$ is a function of $y$. When approximating this situation with actuator disc type
theory, we make the following assumptions:
\begin{enumerate}
  \item The drag in each streamtube, which in the numerical model is applied
    over a length $\Delta x(y)$, is collapsed in the $x$-direction and applied
    at a single point along the streamline, representing an infinitesimal
    actuator disc with cross section $H\dy$.
  \item The results in each of the streamtubes are independent of one another.
    This means that we take no blockage effects into account and assume laminar flow.
\end{enumerate}

For simplicity we first consider a triangle that is oriented in such a way that
it is at its widest at $y=\Delta y$, in other words its top edge is aligned with
the streamline at $y=\Delta y$ (see figure \ref{fig:aligned_triangles}). Furthermore, we have $\Delta x(y=0)=0$ in the
bottom vertex, and $\Delta x(y)$ varies linearly for $0\leq y\leq\Delta y$.
Its area can be computed as 
$A=\tfrac 12\Delta x(\Delta y)\Delta y$. The function $\Delta x(y)$ is therefore
given by:
\begin{equation}
  \Delta x(y) = \frac {2A}{\Delta y^2} y.
  \label{eq:deltax_expression}
\end{equation}
The force applied in each streamtube is given by
\begin{equation}
  \dF = \Delta x(y) \dy \rho c_t u_1(y)^2,
  \label{eq:delta_force}
\end{equation}
where $u_1(y)$ is the velocity through the streamtube.  Similar to
\eqref{eq:modified_thrust_coefficient}, we apply actuator disc theory where we
assume that this force is applied over a cross section $H\dy$ and obtain a
modified thrust coefficient:
\begin{equation}
  \hat{C}_t := \frac{dF}{\tfrac 12 \rho H\dy u_0^2}
    = \frac{2\Delta x(y)dy}{H\dy}\frac{u_1^2}{u_0^2}c_t
\end{equation}
Following the same steps as in equations
\eqref{eq:modified_thrust_coefficient}--\eqref{eq:u1_predict}
we may derive
the following relation between $u_1(y)$ and the upstream velocity $u_0$:
\begin{equation}
  u_1(y) = \frac 1 {1+\tfrac 12 \frac{\Delta x(y)\dy}{H\dy}c_t} u_0
  = \frac 1 {1+\frac{A c_t}{H\Delta y^2} y} u_0.
  \label{eq:u1_variation}
\end{equation}
The varying width $\Delta x(y)$ thus leads to a variation of the velocity
$u_1(y)$ for $0\leq y\leq \Delta y$. In the computer models the accuracy of
this variation is limited by the numerical approximations employed.

\newcommand\umike{u^{\text{\tiny MIKE}}_1}
\newcommand\ufluidity{u^{\text{\tiny Fluidity}}_1}
In MIKE, the underlying discretisation is based on a piecewise-constant velocity
in each cell. To estimate the cell average obtained in the model we therefore
evaluate \eqref{eq:u1_variation} in the centroid at $y=\tfrac
23\Delta y$, which gives:
\begin{equation}
  \umike = \frac 1{1+\tfrac 23\frac{Ac_t}{H\Delta y}} u_0.
  \label{eq:u1_average}
\end{equation}
\newcommand\dmax{\Delta x_{\text{max}}}
For the case where the triangle does not have one of its edges aligned with a
streamline, we may consider splitting the triangle into two triangles that share an edge
that \emph{is} aligned along the streamline (see
figure \ref{fig:aligned_triangles}). The length of this shared edge
is the maximum width $\dmax$ of the triangular drag cell in the streamwise direction.
The area of either of the two triangles that the cell is split into,
can be computed as $A_{1,2}=\tfrac 12\dmax\Delta y_{1,2}$, 
where $\Delta y_{1,2}$ is the height of
either triangle. Therefore for each of the triangles we have $A_{1,2}/\Delta
y_{1,2}=\tfrac
12\dmax$. Thus if we apply the same enhanced friction coefficient $c_t$ in both
triangles, it follows that the estimate \eqref{eq:u1_average}
for the cell average of $\umike$ is the same in both triangles:
\begin{equation}
  \umike = \frac{1}{1+\tfrac 13 \frac\dmax H}u_0.
\end{equation}
Moreover, if we define the overall cross-stream width of the original combined triangle as
$\Delta y=\Delta y_1+\Delta y_2$, we again have $A/\Delta y=\tfrac 12\dmax$.
Thus, in the actual model where the original, non-aligned triangular drag cell
is not split, we can use the same equation \eqref{eq:u1_average} for the
estimated average velocity of the entire cell as we did for the aligned case.
\begin{figure}
  \begin{minipage}{0.45\textwidth}
    \def\svgwidth{\textwidth}
    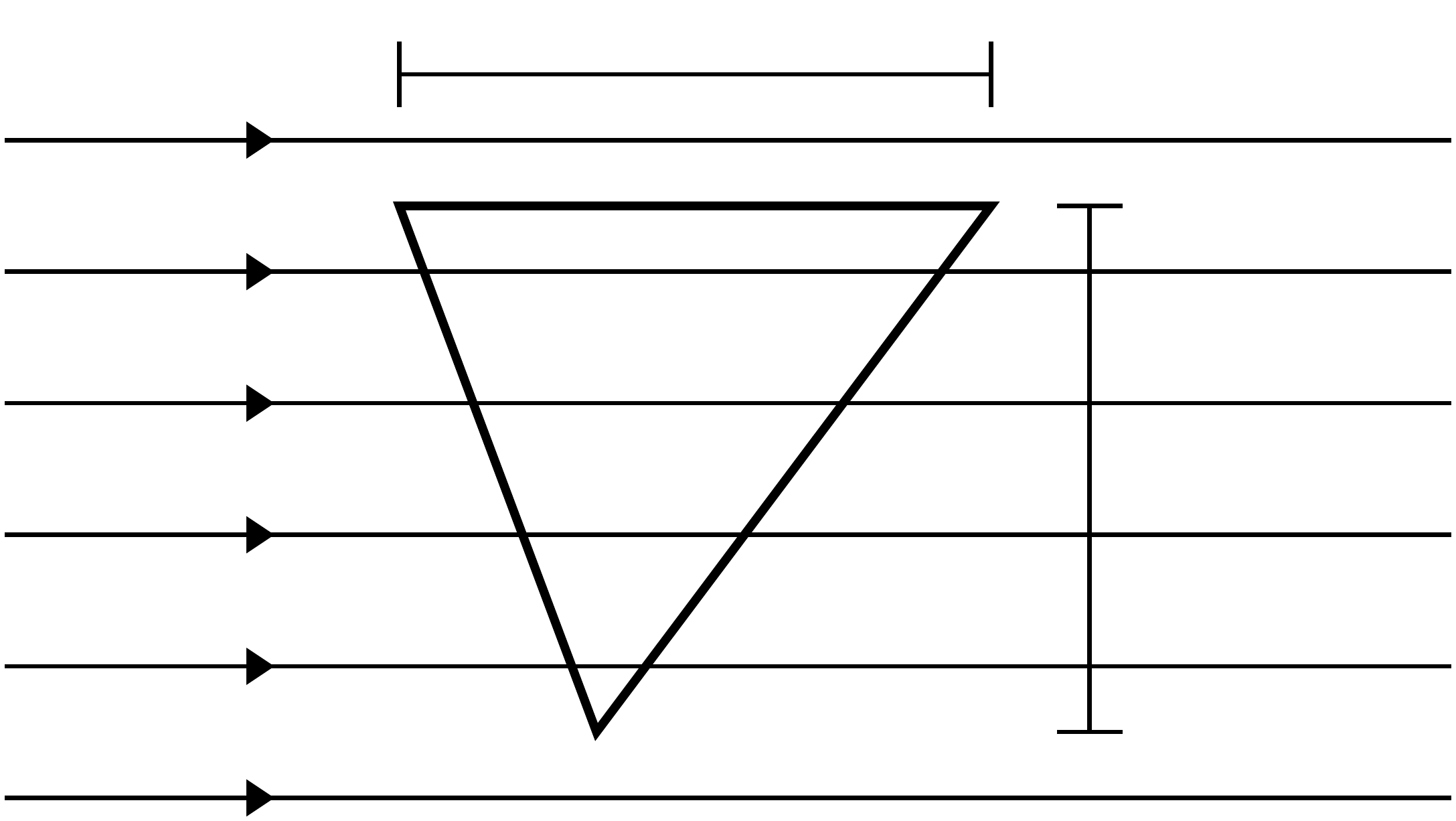
  \end{minipage}%
  \hspace{0.1\textwidth}%
  \begin{minipage}{0.45\textwidth}
    \def\svgwidth{\textwidth}
    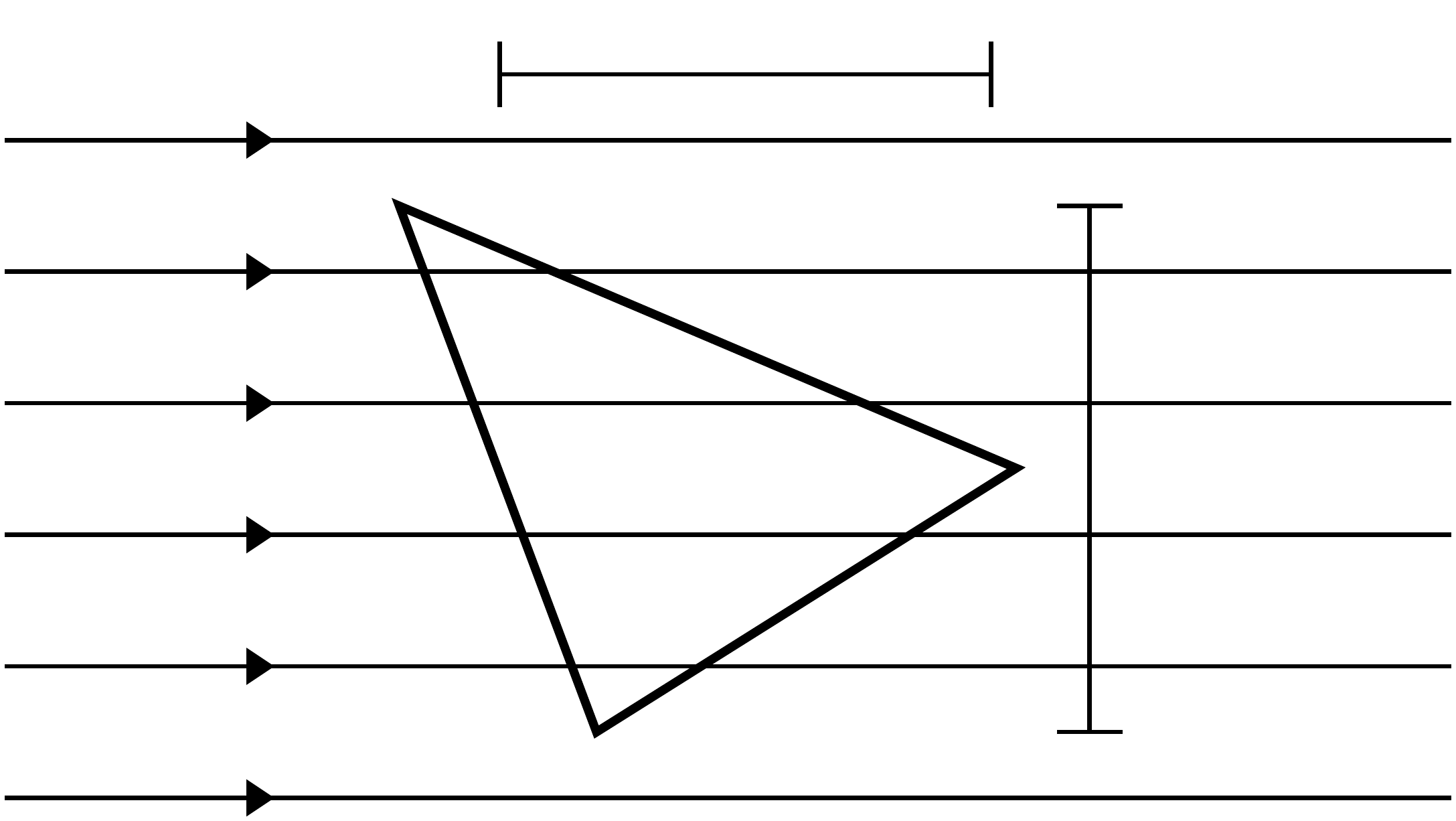
  \end{minipage}
  \caption{Left figure: a triangle with its top edge aligned with the
    streamlines. A coordinate reference frame is chosen, with $0\leq y\leq\Delta
    y$ the coordinate in the cross-stream direction. The width $\Delta x$ 
    of the triangle in the streamwise direction, varies as a function of $y$,
    starting at $\Delta x(y)=0$ at $y=0$, and reaching it maximum width $\Delta
    x(y)=\dmax$ at $y=\Delta y$.
    Right figure: a non-aligned triangle can be divided
    in two triangles that share an edge that \emph{is} aligned with the
    streamlines. In this case, the maximum width $\dmax$ is the length of the shared
    edge.} \label{fig:aligned_triangles}
\end{figure}

Using this estimated average, the force applied in the model is then:
\begin{equation}
  F = A\rho c_t (\umike)^2 = A\rho c_t \left(\frac 1{1+\tfrac 23\frac{Ac_t}{H\Delta
        y}}\right)^2 u_0^2.
\end{equation}
By equating this to the desired force \eqref{eq:quadratic_force}, we may
derive a quadratic expression for $c_t$
\begin{equation}
  -2 A^2 A_t C_t c_t^2 +
  A \left(9 H^2 \Delta y^2 - 6 A_t C_t H \Delta y \right) c_t
  -\tfrac 92 A_t C_t H^2 \Delta y^2 = 0
  \label{eq:mike_correction}
\end{equation}

In Fluidity, the \podg-discretisation prescribes a linear
variation for velocity. Thus we approximate \eqref{eq:u1_variation} by
evaluating it at $y=0$ and $y=\Delta y$ and assuming a linear variation
in between:
\begin{equation}
  \ufluidity(y) = \left(1 - \frac 1{1+\frac{H\Delta y}{Ac_t}} \frac y{\Delta
      y}\right) u_0
  \label{eq:u1_linear}
\end{equation}
The force applied in the model can be found by integrating:
\begin{align}
  F &= \int_{y=0}^{\Delta y} \Delta x(y) \rho c_t (\ufluidity(y))^2 \dy \\
  &= A \rho c_t \left(1 - \tfrac 43
    \left(\frac 1{1+\frac{H\Delta y}{Ac_t}}\right)
    +\tfrac 12
    \left(\frac 1{1+\frac{H\Delta y}{Ac_t}}\right)^2\right) u_0^2.
\end{align}
Equating with the desired force in \eqref{eq:quadratic_force} this time
results in a cubic expression for $c_t$:
\begin{multline}
  A^3 c_t^3 + A^2 \left(4H\Delta y - 3 A_t C_t\right)c_t^2 \\
  + 6A \left(H^2\Delta y^2 -A_tC_tH\Delta y\right) c_t
  - 3A_tC_tH^2\Delta y^2 = 0.
  \label{eq:fluidity_correction}
\end{multline}
In case the triangular drag cell does not have an edge that is aligned with the
streamlines, we may again consider splitting it into two triangles with a shared
edge that is aligned with the flow. Here however, \eqref{eq:u1_linear} does
not predict the same linear function for $\ufluidity$ in both triangles, since 
although $A/\Delta y=\tfrac 12\dmax$ is the same, the value for $\Delta y$ in
the denominator of $y/\Delta y$ is
different for both triangles, and due to the different orientation of the top
triangle, the sign of the gradient of $\ufluidity$ with respect to $y$ will be
opposite. The combined piecewise solution is therefore not supported by the
underlying discretisation. However, we did find that when using the value of $c_t$
found by solving \eqref{eq:fluidity_correction}, the discrete model gave results
that varied only slightly for different orientations of the triangular cell.

\begin{figure}
  \includegraphics[width=\textwidth=0.6\textwidth]{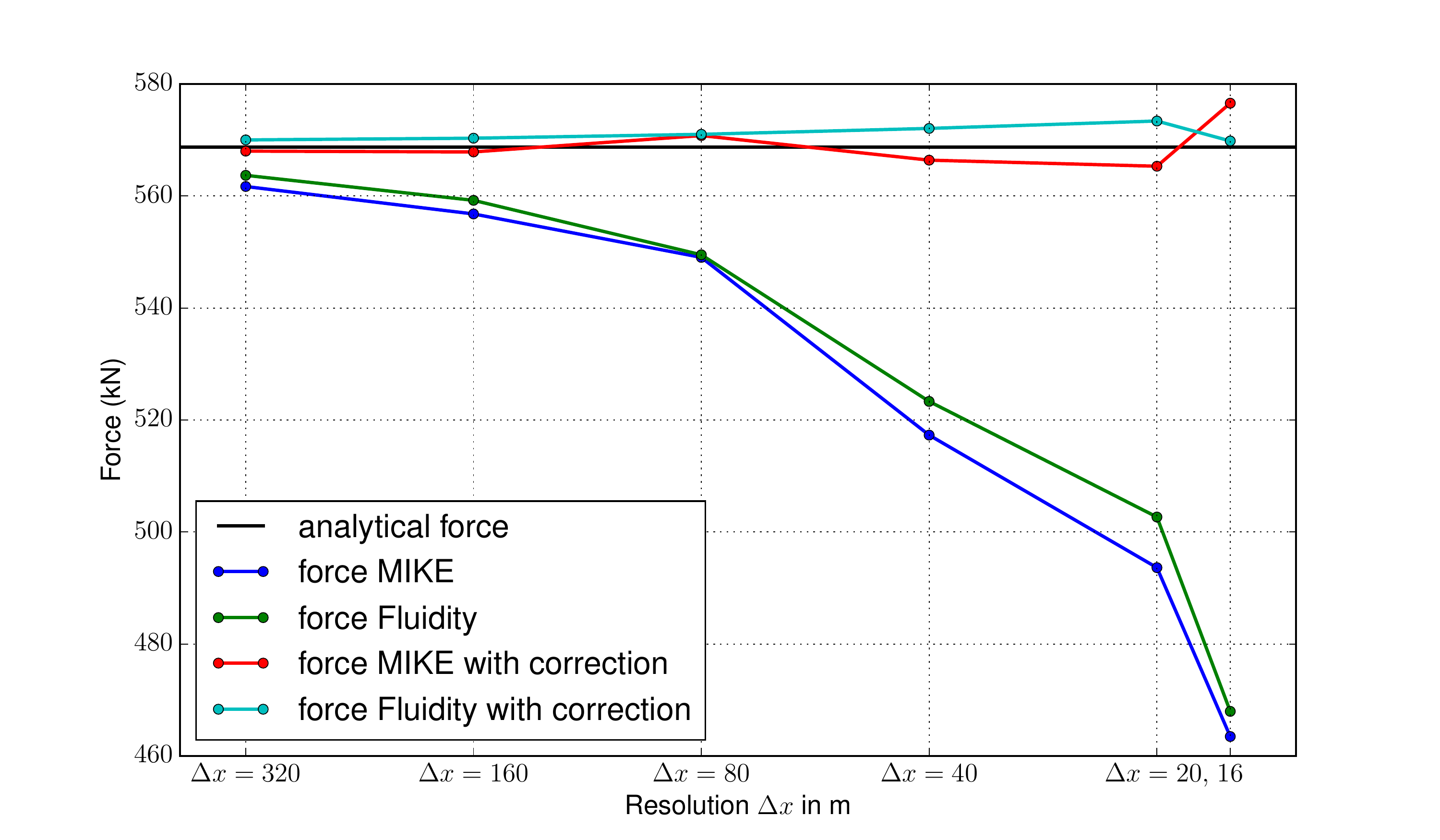}
  \caption{Results for the enhanced drag formulation with the drag applied in 
    a single triangular cell as implemented in both Fluidity and MIKE 21.
    As in figure \ref{fig:decreasing_force_square},
    which show the results for the square case, the force applied decreases
    significantly with increasing mesh resolution. Applying the correction for
    $c_t$ however, given by solving \eqref{eq:mike_correction} (MIKE 21) or
    \eqref{eq:fluidity_correction} (Fluidity), the force
   can be kept more or less constant and much closer to the desired value.}
  \label{fig:decreasing_force_fixed}
\end{figure}
The results in figure \ref{fig:decreasing_force_fixed} indicate that again the
force applied in the unmodified enhanced drag implementation, in
Fluidity and MIKE reduces significantly with increasing mesh resolution. 
A modification to the enhanced bottom drag $c_t$ was derived in this section,
solving for $c_t$ in \eqref{eq:mike_correction} and
\eqref{eq:fluidity_correction} for MIKE and Fluidity respectively,
that is shown here to lead to a force that remains close to the desired value.
The correction in MIKE was implemented by first finding the value for $c_t$ from
\eqref{eq:mike_correction} and then working back from
\eqref{eq:enhanced_drag_coef} to compute what value of $C_t$ should be entered
in the GUI to achieve this value in MIKE.

\section{Power production}
The correction to the enhanced drag formulation, derived in this paper, 
is to ensure that the correct amount of
momentum is extracted from a shallow water model. This means that the force $F$
applied by the enhanced drag in the drag cell (or region) is an accurate
approximation of the real thrust exerted by the turbine on the flow. The amount
of energy taken out of the flow within the cell is given by:
\begin{equation}
  P_{\text{cell}} = F u_{1,\text{model}},
  \label{eq:pmodel}
\end{equation}
where $u_{1,\text{model}}$ is the velocity in the enhanced drag cell. As we
have seen however, in the case where turbines are not fully resolved this velocity
will be larger than the real velocity $u_{1,\text{turbine}}$ that goes through
the turbine (as predicted by actuator disc theory). Therefore, the real power
production $P_{\text{turbine}}=F u_{1,\text{turbine}}$ will be smaller than the
amount of power $P_{\text{cell}}$ taken out of the model in the drag cell.

This discrepancy can be explained from the fact that part of the mixing losses
are not modelled explicitly within the model, but occurs at the sub-grid scale.
Following the analysis of \citet{vogel13}, the total amount of power taken out
of the flow can be split as follows:
\begin{equation}
  P_{\text{total}} = P_{\text{turbine}} + P_{\text{mixing}},
\end{equation}
where $P_{\text{mixing}}$ takes account of the mixing losses due to a.o. shear
between the wake and bypass flows. The total power can be computed as
\citep{vogel13}:
\begin{equation}
  P_{\text{total}} = F u_0.
\end{equation}
Therefore, as long as the model applies an accurate representation 
of the thrust force $F$, using the correction presented in this paper, and an accurate value for the upstream velocity $u_0$,
the total power extracted from the flow in the model will be
accurate as well. The fact that the power $P_{\text{cell}}$
extracted within the drag cell,
according to \eqref{eq:pmodel}, is larger than $P_{\text{turbine}}$ means that
the mixing loss that occurs in the model (outside the drag cell) must be smaller
than the real $P_{\text{mixing}}$ predicted by actuator disc theory. Therefore
part of the mixing loss occurs within the drag cell itself. Thus
$P_{\text{cell}}$ accounts for both the power $P_{\text{turbine}}$
taken out by the turbine itself and additional losses that happen at the
sub-grid level.

\citet{vogel13} considers the case where the drag of an entire farm is smeared
out over an enhanced drag region, with the assumption that all mixing losses
actually occur within this region. In that case it may be assumed that the total
power extraction in the model is a good approximation of the total power
extraction predicted by actuator disc theory, so that the available usefully
extracted power can be computed as a fraction of that using the same theory.

For the case, considered in this paper, where individual turbines are
modelled but are not necessarily fully resolved, part of the mixing losses are
modelled explicitly. As argued above however, using the power extracted from the
flow by the turbine parameterisation still leads to an overprediction of the usefully
extractable energy. It is to be noted that in a shallow water model, even if 
an individual turbine is resolved in the horizontal mesh, with a minimum mesh
distance smaller or equal than the turbine diameter $D$,
the effective cross-section $\hat A_t=\Delta y H$ will still be
larger than the actual turbine cross-section $A_t$. This is because the actual
cross-section does not span the entire depth of the water. Thus, the velocity at
the turbine in the model should be interpreted as a depth-averaged velocity that
averages between the velocity through the turbine, and the bypass velocity
above and below the turbine. This velocity is therefore expected to be higher
than the real turbine velocity itself, and therefore the power extraction by the
depth-averaged turbine-parameterisation will always be an overprediction of the actual
power available to the turbine. The difference between these power values
roughly corresponds to vertical mixing losses that are not explicitly modelled
in the depth-averaged model. In the next section we will explain how
the relationship between the upstream velocity and the local velocity in the
model, derived in this paper, can also be used to predict the usefully
extractable energy, excluding mixing losses, more accurately.

\section{Implementation details}
In this section we summarise, how the analysis derived in this paper can be
practically applied in existing models, in order to ensure that the correct
force is applied on the flow and an accurate estimate of the available turbine
power can be made.

\subsection{Turbine drag applied over a rectangular area}
For models where the turbine parameterisation consists of an enhanced bottom
drag applied over a fixed, rectangular area $A$ (e.g. \cite{vennell10}), we may use the
analysis presented in section \ref{sec:square_cells}. Where existing models typically make
no distinction between upstream and local turbine velocity, they calculate the
enhanced drag coefficient as $c_t=C_t A_t/2 A$. Such implementations can be
improved using the correction given by \eqref{eq:drag_correction_square}. The extra factor at the end
of \eqref{eq:drag_correction_square} can easily be included by the user in either $C_t$ or
$A_t$, without the need for code modification, if these are the input parameters to the model.

An additional complexity arises if $C_t$ itself is not a constant. This
occurs for example if a cut-in speed and/or rating are applied to the turbine. In this case,
$C_t$ is typically given as a function (thrust curve) of the upstream velocity
$u_0$. In the model however only the local velocity $u_1$ is available. Using the formula
\begin{equation}
  u_1 = \tfrac 12\left(1+\sqrt{1-\hat C_t}\right)u_0, \quad
  \hat{C}_t=\frac{A_t}{\hat{A}_t}C_t,
  \label{eq:cell_from_upstream}
\end{equation}
however, it is straight-forward to transform a lookup table that gives the
thrust coefficient for different values of $u_0$, into a lookup table that is a
function of $u_1$, by computing $u_1$ for the given values of $u_0$ as a
pre-processing step.

For the computation of the power available to the turbine, we may use
\eqref{eq:power_formula}. Here, again we use \eqref{eq:cell_from_upstream} to
derive the upstream velocity $u_0$ from the local cell velocity $u_1$. Combining
these two equations, we derive:
\begin{equation}
  P_{\text{turbine}} = 
  \frac{2\left(1+\sqrt{1-C_t}\right)}{\left(1+\sqrt{1-\hat C_t}\right)^3} C_t
  \rho A_t u_1^3.
  \label{eq:extractable_power_square}
\end{equation}
Again, in the case that $C_t$ is not a constant, a lookup table may be used to
obtain the correct value of $P_{\text{turbine}}$ for each value of $u_1$.

\subsection{Turbine parameterisation in an arbitrary triangular mesh}
For models such as MIKE 21 and Fluidity that employ triangular meshes and which implement
turbines through an increased drag applied within a single triangle, the theory
presented in section \ref{sec:triangular_cells} can be applied. In triangular mesh models
where the drag force is based on a cell-averaged velocity, the value
for the enhanced drag coefficient can be found by solving 
\eqref{eq:mike_correction} for $c_t$. Models that use a linear interpolation of
velocities stored in the vertices, such as Fluidity 
should use the value of $c_t$ found by solving \eqref{eq:fluidity_correction}.
The same approach could also be followed to implement a turbine in a single drag
cell in Telemac 2D, where its Finite Element modus is expected to behave in a similar manner as
Fluidity, using a linear representation of the velocity within a cell.

In models, like MIKE, where the applied drag force and the associated
coefficient $c_t$ are not explicitly prescribed, the same effect can be achieved by
modifying the value of $C_t$. This is done by assuming the implementation
is equivalent to the standard enhanced bottom drag formulation according to
equation \eqref{eq:enhanced_drag_coef}. Indeed the results in figure
\ref{fig:decreasing_figure} where the standard drag implementation of Fluidity
is compared with results in MIKE show that this is true to at least a good
approximation. By providing MIKE with a modified value of $C_t$
\begin{equation}
  C_{t, \text{modified}} = \frac{2 A c_t}{A_t},
\end{equation}
we can therefore create the effect of applying a value of $c_t$ obtained from
\eqref{eq:mike_correction} without modifying the code. Note, that in equation
\eqref{eq:mike_correction}  we use the original value
of $C_t$ for the real turbine.

For non-constant $C_t$ that is given as a thrust curve, MIKE (and similar
models) use the local cell velocity $u_1$ instead of the upstream velocity to
look up the value of $C_t$. This can be corrected by converting the upstream values
$u_0$ in a $u_0\to C_t$ look-up table into cell velocities $u_1$ using equation
\eqref{eq:u1_average}.

To compute the power that can be usefully extracted by the turbine we again
use \eqref{eq:power_formula} this time combined with \eqref{eq:u1_average},
giving:
\begin{equation}
    P_{\text{turbine}} = \tfrac 14 \left(1+\sqrt{1-C_t}\right) C_t A_t
      \left(1+\tfrac 23 \frac{Ac_t}{H\Delta y}\right)^3
      (\umike)^3.
  \label{eq:extractable_power_mike}
\end{equation}

For finite element models, such as Fluidity, that consider a linear variation of
the velocity within the cell we can use \eqref{eq:u1_linear} which predicts the
relationship between the upstream velocity and the velocity in the cell as a function of $y$.
By first taking an average of the finite element solution $\ufluidity$ within
the drag cell in the streamwise direction ($x$-direction), we can then use 
this equation to estimate the upstream velocity $u_0$. This estimate may in
practice still vary in the cross-streamwise direction ($y$-direction), so we take the
cell average of its cube to obtain an estimate for $u_0^3$ in
\eqref{eq:power_formula}. Combining all this gives:
\begin{align}
  P_{\text{turbine}} 
    &= 
       \tfrac 14\left(1+\sqrt{1-C_t}\right)\rho\frac{C_tA_t}A
       \int_{y=0}^{\Delta y} \Delta x(y) \left(
         \frac{\frac{\int_{x=0}^{\Delta x(y)}\ufluidity dx}{\Delta x(y)}}{1-\frac{1+H\Delta y}{Ac_t}\frac y{\Delta
             y}}\right)^3 dy.
\end{align}

\subsection{Support structure}
The drag exerted on the flow by the support structure, e.g. pylons or tripods,
can typically also be parameterised as a force that depends quadratically on the
upstream velocity $u_0$:
\begin{equation}
  F_{\text{support}} = \tfrac 12 \rho C_s A_s u_0^2,
\end{equation}
where $A_s$ and $C_s$ are the cross-sectional area and the drag coefficient of
the support structure.
To include this drag in the form of an enhanced bottom drag coefficient,
we have to deal with the same issue of expressing this force in terms
of a local velocity $u_1$. Although the theory derived so far can be
straightforwardly applied to any force in this quadratic form, we cannot 
simply derive the enhanced bottom drag coefficient that represents the support
drag, denoted by $c_s$, independently of $c_t$ and add them up.
This is because the local velocity $u_1$ is the depth-averaged velocity that
will be slowed down by the two sources of drag simultaneously.

For the drag parameterisation applied over a square area, the correct value for
$u_1$ is obtained from \eqref{eq:cell_from_upstream} by using
\begin{equation}
  \hat{C_t} = \frac{A_tC_t + A_sC_s}{H\Delta y}.
  \label{eq:combined_ct}
\end{equation}
We can then derive a combined enhanced drag coefficient 
\begin{equation}
  c_{t} = \frac{A_tC_t + A_sC_s}{2A} 
    \frac 4{\left(1+\sqrt{1-\hat{C_t}}\right)^2},
\end{equation}
that represents both turbine and support drag (cf.
equation \eqref{eq:drag_correction_square}). The effectively extracted power
is still given by \eqref{eq:extractable_power_square} using the combined value
of $\hat{C_t}$ from \eqref{eq:combined_ct}, but only using the values for the
turbine itself for $C_t$ and
$A_t$.

For models where the drag is applied over a triangular cell, the relation
between $u_0$ and $u_1$ is expressed in terms of the actual enhanced bottom drag
coefficient $c_t$. Thus if we include both turbine and support drag in $c_t$, we
can maintain equations \eqref{eq:u1_average} and \eqref{eq:u1_linear} for models
with cell-wise constant, and piecewise linear velocities respectively. The
actual combined value of $c_t$ can then be found from the quadratic and cubic
equations \eqref{eq:mike_correction} and \eqref{eq:fluidity_correction}
respectively, by replacing $A_tC_t$ with $A_tC_t+A_sC_s$. Finally, the
extracted power is still given by \eqref{eq:extractable_power_mike}, using only
the turbine values for $C_t$ and $A_t$, but using the combined value of $c_t$.

\section{Conclusions}
In order to accurately estimate the resource available to tidal turbines and to
assess their impact on the hydrodynamics, it is important to accurately
represent the drag force exerted by the turbines on the flow. In depth-averaged,
and more generally under-resolved hydrodynamic models, one should keep in mind
that the local model velocity at the turbine is different from both the upstream
and the actual velocity passing through the turbine. The relationship between them
is dependent on the mesh resolution, and in the case of depth-averaging, the ratio
between the actual turbine cross section and the flow cross section spanning the
entire depth. Therefore, although the use of the local velocity for the
implementation of the drag force is computationally attractive, it is required
to take these relationships into account to avoid spurious and mesh-dependent
results. In addition, a better understanding of
the relation between local and upstream velocity is necessary for an
accurate estimate of the power available to the turbine.

Here we have presented the theory for a single, isolated turbine, and
demonstrated that a correction based on linear momentum actuator disc 
theory taking into account the actual numerical cross section that the force is
applied over in the model, can be used to obtain results that are consistent
over a range of grid scales. It was shown that the standard
enhanced bottom drag formulation results in a drag force that decreases with
decreasing grid lengths, in particular when the grid size falls below the length
scale of the turbine wake (roughly 10--20 turbine diameters). With the correction
the applied force can be kept constant to a large degree, thus ensuring that the
effect of the turbine on the large scale flow is correctly modelled.

The analysis for single, isolated turbines 
may be sufficient for sparsely populated turbine sites which see little
interaction between turbines. It is generally recognised however, that in
order to achieve the maximum available energy from certain sites, one needs to
consider turbine configurations that benefit from local and global blockage
effects \citep{nishino12}, e.g. fence structures. The analysis in this paper could be extended to
include blockage effects. Here again one should make a distinction between the
influence of blockage on the relation between upstream and turbine velocities in
reality, and the influence of blockage on the relation between upstream and local
velocities in the model, in particular taking into account the difference in
effective cross sections between reality and the model. 

With more closely packed turbines the representation of turbine wake structures and wake 
recovery also becomes much more important. In addition, the turbulence
characteristics may have a great impact on the performance of the turbines. 
As mentioned in the introduction, depth-averaged models will not be sufficient
to accurately model these three-dimensional near-field effects. In further work
we would like to explore however, how well these effects can still be
approximated in depth-averaged models, possibly through parameterisation and
tuning of horizontal turbulence models. Nonetheless, we recognise
that in general it may no longer be possible to simply
extrapolate from the results of a single isolated turbine, and it may be
required to study the effects of combining multiple turbines in detailed
three-dimensional CFD calculations and lab experiments.

\section{Acknowledgements}
The authors would like to kindly acknowledge the UK's Engineering and Physical
Sciences Research Council (EPSRC), projects EP/J010065/1 and EP/M011054/1, for funding which supported this work. Part of the computations in this work have been made possible using the Imperial College High Performance Computing Service.

\appendix
\section{Linear Momentum Actuator Disc Theory}
\label{sec:lmadt}
In this appendix we briefly review the main steps in the derivation of the
actuator disc theory used in tidal turbine calculations. This is so we can refer
to the relevant equations when the modifications, that take into account the
numerical implementation details of the enhanced bottom drag formulation, 
are derived in the main text. These results can be found in e.g.
\citet{garrett07}, or \citet{whelan09}.

We consider a channel of cross-sectional area $A_c$ in which a turbine is located
with cross section $A_t$. We assume a uniform flow across the channel upstream
of the turbine with velocity $u_0$, the flow through the turbine is $u_1$.
Further downstream we define $u_3$ to be the velocity in the wake, and $u_4$
the bypass velocity. Furthermore we assume that at the point down-stream where
$u_3$ and $u_4$ are defined we have a uniform water level $\eta_4$. The water
level upstream is denoted by $\eta_0$, and the water levels just upstream and
downstream of the turbine, associated with the pressure drop across the turbine
are denoted by $\eta_1$ and $\eta_2$.

First we formulate the conservation of mass for the flow through the turbine and
in the bypass flow
\begin{align}
  A_t u_1 &= A_3 u_3, \label{continuity1} \\
  A_c u_0 &= A_3 u_3 + (A_c-A_3) u_4 \label{continuity2},
\end{align}
where $A_3$ is the cross-sectional area of the wake at the location where $u_3$ is
defined. Here we neglect the influence of the water level on the cross sections,
so that the cross-sectional area of the bypass flow is given by $A_c-A_3$. Inclusion
of the dependency of cross section on the water level is only significant for
high Froude numbers, with details given in \citep{whelan09}.

The force $F$ exerted by the turbine on the flow (and vice-versa),
can be related to a conservation of momentum principle in the entire channel, or
to the pressure drop across the turbine:
\begin{align}
  F &= A_c \rho u_0^2 - A_3 \rho u_3^2 - (A_c-A_3) \rho u_4^2 + \rho g A_c (\eta_0-\eta_4),
  \label{force1} \\
  F &= \rho g A_t (\eta_1-\eta_2), \label{force2}
\end{align}
where $g$ is the gravitational acceleration. Finally, 
applying Bernoulli's principle along streamlines: 1) from upstream, where $u_0$
is considered uniform, to just before the turbine, where water level $\eta_1$ is
defined; 2) from just after the turbine, where water level $\eta_2$ is defined,
to downstream where a uniform water level $\eta_4$ is defined; and 3) in the
bypass flow from upstream to downstream.
This yields three more equations:
\begin{align}
  \tfrac 12 u_0^2 + g\eta_0 &= \tfrac 12 u_1^2 + g\eta_1, \label{bern1} \\
  \tfrac 12 u_1^2 + g\eta_2 &= \tfrac 12 u_3^2 + g\eta_4, \label{bern2} \\
  \tfrac 12 u_0^2 + g\eta_0 &= \tfrac 12 u_4^2 + g\eta_4. \label{bern3}
\end{align}
Assuming boundary conditions for $u_0$ and $\eta_4$, and an expression for $F$ as
a function of $u_0$, we have seven equations 
for seven unknowns: $u_1, u_3, u_4, \eta_0, \eta_1, \eta_2,$ and $A_3$.

\subsection*{General solutions}
The Bernoulli equations \eqref{bern1} to \eqref{bern3} can be rewritten as
expressions for water level differences:
\begin{align}
  g\eta_1 - g\eta_2 &= g\eta_0 - g\eta_4 + \tfrac 12 \left(u_0^2 - u_3^2\right), \label{bern4} \\
  g\eta_0 - g\eta_4 &= \tfrac 12 \left(u_4^2 - u_0^2\right), \label{bern5}
\end{align}
and thus
\begin{equation}
  g\eta_1 - g\eta_2 = \tfrac 12\left(u_4^2-u_3^2\right). \label{bern6}
\end{equation}
We can therefore rewrite the two expressions \eqref{force1} and
\eqref{force2} as:
\begin{align}
  F &= A_3 \rho\left(u_4^2-u_3^2\right)
  - \tfrac 12 A_c\rho\left(u_4^2-u_0^2\right) 
  \label{forcer1} \\
  F &= \tfrac 12 A_t\rho\left(u_4^2-u_3^2\right) \label{forcer2}
\end{align}
Equations \eqref{continuity2}, \eqref{forcer1} and \eqref{forcer2} give three
equations for the three unknowns $u_3, u_4$ and $A_3$. 
Substitution of $A_3(u_4-u_3)=A_c(u_4-u_0)$ from \eqref{continuity2}, in \eqref{forcer1} eliminates $A_3$:
\begin{equation}
  \begin{split}
  F &= A_3 \rho\left(u_4-u_3\right)(u_4+u_3) 
  - \tfrac 12 A_c\rho\left(u_4^2-u_0^2\right) \\ 
  &= A_c \rho\left(u_4-u_0\right)(u_4+u_3) 
  -\tfrac 12 A_c\rho\left(u_4^2-u_0^2\right) \\
  &= A_c\rho\left(u_4-u_0\right)\left(u_3+\tfrac 12 u_4 - \tfrac 12 u_0\right).
\end{split}
  \label{forcer3}
\end{equation}
We can rearrange \eqref{forcer2} and \eqref{forcer3} in the following manner,
respectively:
\begin{align}
  A_c^2 \left(u_4-u_0\right)^2 u_3^2 &= 
  A_c^2 \left(u_4-u_0\right)^2\left(u_4^2 - \frac {2F}{A_t\rho}\right), 
     \label{u3expression1} \\
  A_c^2 \left(u_4-u0\right)^2 u_3^2 &= 
     \left(F-\tfrac 12 A_c\rho\left(u_4-u_0\right)^2\right)^2.
     \label{u3expression2}
\end{align}
We introduce the additional definitions,
\begin{equation}
  C_t := \frac{F}{\tfrac 12 A_t \rho u_0^2},\quad
  \text{ and }\quad
  \epsilon := \frac{A_t}{A_c}.
  \label{ct_definition}
\end{equation}
Note that we do not have to assume that $F$ is actually quadratic in $u_0$, so that
$C_t$ is not necessarily a constant; it may still be
dependent on $u_0$. With these we can derive the following quartic 
polynomial in $u_4$ from \eqref{u3expression1} and \eqref{u3expression2}:
\begin{equation}
  \frac{1}{4} \left(C_{t} \epsilon -
  \left(\frac{u_{4}}{u_{0}} -1\right)^{2}\right)^{2}
  -\left(\frac{u_{4}}{u_{0}}-1\right)^{2} 
  \left( \frac{u_{4}^{2}}{u_{0}^{2}} - C_{t} \right) = 0.
  \label{polynomialu4}
\end{equation}
Finally, by \eqref{forcer2}:
\begin{equation}
  u_3 = \sqrt{u_4^2 -C_t u_0^2}, \label{u3fromu4}
\end{equation}
and $A_3$ can be derived by again substituting \eqref{continuity2} in
\eqref{forcer1} but this time to eliminate $A_c(u_4-u0)$, so that
\begin{equation}
  F = A_3 \rho\left(u_4-u_3\right)\left(u_3+\tfrac 12 u_4 -\tfrac 12 u_0\right),
\end{equation}
which in combination with \eqref{forcer2}, gives:
\begin{equation}
  A_3 = \frac{\tfrac 12 u_4+\tfrac 12 u_3}{u_3+\tfrac 12 u_4 - \tfrac 12 u_0}A_t.
  \label{A3expression}
\end{equation}

\subsection*{Zero blockage limit}
From the above, it follows that in the limit $\epsilon\to 0$: $u_4\to
u_0$ and thus $\eta_4\to \eta_0$. In this limit, \eqref{u3fromu4} becomes
\begin{equation}
  u_3 \to \sqrt{1 - C_t} ~u_0,
\end{equation}
and combining \eqref{A3expression} and \eqref{continuity1}:
\begin{equation}
  u_1 = \frac{\tfrac 12 u_4+\tfrac 12 u_3}{u_3+\tfrac 12 u_4 - \tfrac 12 u_0}
  u_3 \to \tfrac 12 \left(1+\sqrt{1-C_t}\right) u_0.
  \label{u1fromu0}
\end{equation}
The energy yield then becomes:
\begin{equation}
  P = F u_1 \to \tfrac 14 \left(1+\sqrt{1-C_t}\right)C_t A_t \rho u_0^3.
  \label{eq:power_formula}
\end{equation}
The maximum yield as a function of $C_t$ is obtained by:
\begin{gather}
  \frac{d}{dC_t} \left[\left(1+\sqrt{1-C_t}\right)C_t\right] =
  \frac{1-\tfrac 32 C_t+\sqrt{1-C_t}}{\sqrt{1-C_t}} =0 \\
  \implies \left(\tfrac 32 C_t-1\right)^2 = 1-C_t \implies C_t=\frac 89.
\end{gather}
Thus the maximum power (assuming no blockage) is
\begin{equation}
  P_{\text{max}} = \frac{16}{27} \cdot \tfrac 12 A_t\rho u_0^3 \approx 0.59 \cdot
  \tfrac 12 A_t\rho u_0^3.
  \quad \text{({\bf Betz limit})}
\end{equation}

\bibliographystyle{elsarticle-harv}
\bibliography{references}
\end{document}